\documentstyle[11pt,amssym]{article}
\begin{document}
\newcommand{\xV}{\vec{x}}
\newcommand{\tO}{\tau_{0}}
\newcommand{\AT}{A_{\rm T}}
\newcommand{\oS}{q_{S}}
\newcommand{\tS}{\tau_{S}}
\newcommand{\tL}{\tau_{L}}
\newcommand{\kV}{\vec{k}}
\newcommand{\Mtil}{\widetilde{M}}
\newcommand{\Hpres}{\mbox{$100h\,{\rm km\,sec^{-1}\,Mpc^{-1}}$}}
\renewcommand{\thesection}{\Roman{section}}
\begin{center}
{\Large \bf LIMITS ON A STOCHASTIC BACKGROUND OF GRAVITATIONAL 
WAVES FROM GRAVITATIONAL LENSING}
\end{center}

\begin{center}
Rennan Bar-Kana\footnote{Electronic address: barkana@arcturus.mit.edu}

{\it Department of Physics, MIT, Cambridge, MA 02139}

(Received 10 June 1996; revised manuscript received 6 August 1996)
\end{center}

\vspace{.2in}

\begin{quote}
\begin{center} {\bf Abstract:} \end{center}
We compute the effects of a stochastic background of gravitational
waves on multiply imaged systems or on weak lensing. There are two
possible observable effects, a static relative deflection of images
or shear, and an induced time dependent shift or proper motion. We
evaluate the rms magnitude of these effects for a COBE normalized,
scale-invariant spectrum, which is an upper limit on spectra
produced by inflation. Previous work has shown that large-scale
structure may cause a relative deflection large enough to affect
observations, but we find that the corresponding effect of gravity
waves is smaller by $\sim 10^4$ and so cannot be observed. This
results from the oscillation in time as well as the redshifting of
the amplitude of gravity waves. We estimate the magnitude of the
proper motion induced by deflection of light due to large-scale
structure, and find it to be $\sim 10^{-8}$ arcsec per year. This
corresponds to $\sim 50$ km/s at cosmological distances, which is
quite small compared to typical peculiar velocities. The COBE
normalized gravity wave spectrum produces motions smaller still by
$\sim 10^2$. We conclude that light deflection due to these
cosmological perturbations cannot produce observable proper motions
of lensed images. On the other hand, there are only a few known
observational limits on a stochastic background of gravity waves at
shorter, astrophysical wavelengths. We calculate the expected
magnitudes of the effects of lensing by gravity waves of such wavelengths, 
and find that they are too small to yield interesting limits
on the energy density of gravity waves.
\end{quote}

\vspace{.2in}

\section{Introduction}

Events in the early Universe may have left a stochastic
background of gravitational waves (GW's). In particular, a
generic prediction of inflation is a relic spectrum of
GW's \cite{AbbWs}. Detecting these elusive remnants 
would not only establish this prediction of general
relativity, but also serve as a critical test for
inflation. While the predicted background may be
too weak for direct detection \cite{me}, it could be
detected indirectly through its effect on light
propagation in the Universe. Even if the effects of GW's cannot 
be distinguished observationally from other effects,
observers who assume no GW's might reach incorrect conclusions about 
the distribution of matter in the Universe.

Gravitational lensing is one of the most promising methods
of mapping the distribution of matter at cosmological distances.
Detailed observations of multiple images of quasars have been
used to try to reconstruct the lensing mass distribution (e.g.,
\cite{falco}). It has also long been recognized that
measurements of the time delay between images can be used to
determine the Hubble constant \cite{refsdal}.
Gravitational lenses and sources, however, typically lie at
significant redshifts. Light rays are thus deflected by
large-scale structure (LSS) and GW's as they traverse the 
cosmological distance to the observer, and these deflections may 
change the simple lensing picture.

GW's may be produced by many sources. Astrophysical sources, such
as close binary systems which include a neutron star or black
hole, radiate GW's, and numerous individual sources may
superpose to create a stochastic background. At the Planck time, 
quantum fluctuations in the metric are significant and may produce 
gravitons. Phase transitions in the universe may lead to topological
defects such as cosmic strings, which generate GW's. A period of 
inflation may leave behind a significant amount of GW's. Whatever the 
source, any spectrum which extends over wavelengths comparable to 
the present horizon
would contribute to the quadrupole anisotropy of the
cosmic microwave background (CMB) \cite{infl}. Such a spectrum
is therefore limited by the anisotropy measured by the
COBE DMR experiment \cite{COBE}. For our calculations
we adopt a scale-invariant primordial spectrum, i.e., one 
which has constant energy density per logarithmic frequency,
which we assume produces the entire measured quadrupole 
anisotropy. Inflationary models predict slightly 
tilted spectra which are responsible only for some fraction
of the anisotropy \cite{infl,me}, and so are generally 
weaker than our adopted case.

In inflation, GW's are produced in conjunction
with density fluctuations. The initial nearly scale-invariant power 
spectrum of density fluctuations evolves as modes reenter
the horizon after inflation, and as
structure later forms in a universe dominated by dark matter.
The present spectrum is strongly constrained by galaxy and cluster 
surveys, and can be used to study the effects of LSS
on lensing. The induced effects are small but potentially
observable. In weak lensing, the effect is a coherent distortion
of background galaxies by an ellipticity of the order of a few per cent 
\cite{weak,kaiser}. In strong lensing, the primary effect is an 
external shear which may be 
significant for observed four-image systems \cite{me2,seljak}. 

In general, the influence at a given time of a weak metric 
perturbation on 
light propagation is simply described by two effects. Their 
magnitudes were 
estimated for LSS in Ref.\ \cite{seljak}, which we summarize here. 
The first effect is a constant deflection, the same for all nearby 
light rays. 
This deflection simply displaces the ``true'' angular position of 
an observed lens 
or source, and is not directly observable. In the case of LSS, 
deflections from coherent structures of size $\sim 1$ Mpc 
add up in a random walk, 
giving an overall deflection of order a few arcminutes at redshift 
$1$, which scales as the square root of comoving distance $r$. The 
second effect 
is a {\it relative} deflection between nearby light rays, which 
produces a 
focusing and shear with observable effects on weak and strong 
lensing.
For two rays at initial angle $\theta$, each coherent structure at 
a distance
$r$ causes a relative deflection proportional to their separation 
of $\approx r \theta$. The additional random walk gives a relative 
angular fluctuation 
of $\approx 0.07\ \theta$ at redshift $1$, which scales as 
$r^{3/2}$. 

It was suggested in Ref.\ \cite{allen} that gravity waves could 
significantly
affect the time delays in a multiply imaged system. It was later 
pointed
out \cite{surpi} that a correct analysis must include the lensing 
constraint,
i.e., the fact that image rays in the presence of GW's follow 
different
paths than for no GW's, so that all rays go from the source to a 
common destination, the observer. These later authors also 
showed that both LSS and GW's have no observable effects on lensing, 
to lowest order. However, in their lowest order expansion they 
assumed that two image rays that are observed at an angular 
separation $\theta$
are separated by a distance of exactly $r\theta$ on the lens plane 
at a distance $r$. In other words, they neglected the relative 
deflection between light rays, and therefore only included an 
overall, constant deflection due to LSS or GW's.

We can easily see why this assumption leads to no observable 
effects. In the absence of metric perturbations, we can write the 
lens equation for a thin lens as (e.g., \cite{schneider}) 
$\vec{\beta}=\vec{\theta}-\vec{\alpha}_{\rm lens}
(\vec{\theta})$, where $\vec{\theta}$ and $\vec{\beta}$ are the 
image and source angles, respectively, and $\vec{\alpha}_{\rm lens}$ 
is the scaled deflection
angle, which is determined by the mass distribution of the lens. 
If we neglect relative deflections, then LSS or GW's can only cause 
an angular shift $\vec{\alpha}_{L}$ between the observer and the 
lens, and a shift $\vec{\alpha}_{LS}$ between the lens and the 
source. Then the
lens equation becomes $\vec{\beta}=\vec{\theta}-\vec{\alpha}_{\rm lens}
(\vec{\theta})+\vec{\alpha}_{S}$, where $\vec{\theta}$ is now 
measured relative to the {\it observed} (and shifted) lens position,
and $\vec{\alpha}_S$ involves $\vec{\alpha}_{L}$ and $\vec{\alpha}_
{LS}$ (see Sec.\ III for the full details). 
The constant (i.e., $\vec{\theta}$-independent) deflection 
$\vec{\alpha}_{S}$ has no effect on any observables of the lens system 
(e.g., \cite{schneider}), since 
$\vec{\beta}$ is not directly observable . Fermat's principle then 
implies that
the lens equation must be equivalent to $\partial \Delta t/\partial 
\vec{\theta}
=0$ at fixed $\vec{\beta}$, where $\Delta t$ is the relative time 
delay. There is 
thus no observable effect on the time delay, either, since it can 
be derived from the lens equation, up to (unobservable) 
$\vec{\theta}$-independent terms.

This approximation of neglecting the relative deflection may not be
a good one. Indeed, such deflection can have observational 
consequences, which may be sufficiently large to detect in the case
of LSS \cite{me2,seljak}. In this paper, we compute the rms total 
and relative deflections between light rays induced by a 
scale-invariant stochastic background
of GW's. Unlike LSS, GW's oscillate with time, and so the effect of short 
wavelength modes does not amplify, as light rays deflect one way 
in crests and the 
opposite way in troughs. In addition, the energy density and thus 
also the amplitude
of sub-horizon GW's redshift away as the universe expands. The lensing
effect is thus 
dominated by wavelengths on the scale of the distance to the source.
Each such mode 
acts as a single coherent structure, and so both the 
total and relative deflections due to GW's scale approximately
linearly with distance. The effect of different modes must be
convolved with a particular power spectrum and include the 
above-mentioned 
decay of each mode as the universe expands. We find simple
integral expressions for the scale-invariant spectrum. The
total and relative deflections are smaller than those caused by
large-scale structure by factors of the order of $10^{2}$ and $10^{4},$ 
respectively. We do not need to explicitly set up the lens equation,
since the rms shear in the lens equation is directly related to
the rms relative deflection of light rays, which we calculate.
This fact was demonstrated for LSS in Ref.\ \cite{me2}, and we give 
a general proof in Sec.\ III below. Our results imply that the static 
effects of the GW spectrum on lensing are negligible compared to 
those of LSS, and cannot be detected in practice. 

In addition to the static effects of LSS and GW's on lensing, it
is possible that the fluctuation in the induced deflection with time
would be directly manifested as an observed proper motion of images.
In other words, the sources do not really move but the light rays 
from the sources are deflected and so the sources appear to move. 
We find that even LSS can only produce motions of order $10^{-8}$ 
arcsec per year from this effect.
This corresponds to $\sim 50$ km/s at a distance of a Gpc,
and the effect of GW's is smaller still by a factor of $\sim 10^{2}$. 
Since typical peculiar velocities are much larger, the proper motion
induced by deflection of light due to LSS is unobservable, and the 
same is true for the COBE-normalized scale-invariant spectrum of GW's.

However, we may try to use shear or proper motions of imaged sources 
to improve 
existing limits on stochastic GW's at a range of astrophysical 
wavelengths. There are only a few such limits known:
Single-pulsar timing yields $\Omega_{\lambda}<1 \times 10^{-8}$ at 
$\lambda
\approx 2$ pc \cite{latest,kaspi}, binary pulsar timing implies 
$\Omega_{\lambda}< 0.04$ over $\lambda \approx 2$ pc to 1 kpc and
$\Omega_{\lambda}< 0.5$ up to $10$ kpc \cite{latest,Rees}, and 
the observed angular correlation function of galaxies sets a limit 
of $\Omega_{\lambda}< 
10^{-3}$ over $\lambda \approx 
100$ kpc-$100$ Mpc \cite{linder}. These limits apply to any 
stochastic background of GW's, 
whether cosmological in origin or generated at low redshift as a 
superposition of many discrete sources.
For a cosmological spectrum that existed at early times, there are 
also big bang 
nucleosynthesis constraints of $\Omega_{\lambda}< 10^{-4}$ for 
$\lambda < 100$ pc 
\cite{Carr} and CMB limits of $\Omega_{\lambda}< 10^{-12}$ at 
Horizon wavelengths (from COBE) and $\Omega_{\lambda}< 10^{-8}$ 
for $\lambda > 1$ Mpc from small-scale anisotropy \cite{linder2}. 

In Ref.\ \cite{linder} it was suggested that highly magnified lensed
sources could increase the sensitivity to detecting proper motions 
due to GW's. The angular deviations 
induced by GW's produced by an individual source were discussed in 
Ref.\ \cite{fakir}. Ref.\ \cite{pyne} considered detecting proper 
motions (of unlensed sources) due to GW's through VLBI measurements, 
but our approach is simpler than theirs.
For an image of a lensed source, only an angular deflection of 
the source {\it relative} to the lens is easily observed, and we
find that this relative motion is small when we assume an
isotropic GW background. Thus we do not find an
interesting limit on the energy density.

\section{Formalism}

In this section we review the formalism describing gravity waves,
their cosmological evolution, and their effect on lensing, as
well as the usual formalism of gravitational lensing.
We work in the framework of a flat Robertson-Walker metric
with small-amplitude tensor metric fluctuations. For weak
perturbations, we can consider the effect of GW's without including
LSS, since the cross terms between them would be of higher order.
In comoving coordinates we can write the line element as
\begin{equation}
ds^2=a^2(\tau)[-d\tau^2+(\delta_{ij}+h_{ij}) dx^{i} dx^{j}]\ .
\label{metric}
\end{equation}
Here $\tau$ is the conformal time, $a(\tau)$ the expansion factor, 
and we have set $c=1$. We expand the metric perturbation in plane 
waves ($k=2\pi / \lambda$),
\begin{equation}
h_{lm}(\xV,\tau)=\int d^{3}k~ h^{n}(\kV,\tau)\epsilon_{lm}^{n}
(\widehat{k})\,e^{-i\kV \cdot \xV},
\end{equation}
where $\epsilon_{lm}^{n}$ is the polarization tensor which depends 
on the direction $\widehat{k}$ ($l$ and $m$ are
spatial indices ranging from 1 to 3, while $n$ ranges over the 
polarization components $+,\times$). For a wave
propagating in the $z$-direction, the nonvanishing components
are in the $x$-and $y$- rows and columns: $$\epsilon_{lm}^{+}= 
\left( \begin{array}
{rrr} 1 & 0 & 0 \\ 0 & -1 & 0 \\ 0 & 0 & 0 \end{array} \right),\ 
\epsilon_{lm}^{\times}=\left( \begin{array}{rrr} 0 & 1 & 0 \\ 1 & 0 & 0
\\ 0 & 0 & 0 \end{array} \right)\ .$$ For
other propagation directions $\kV$, we rotate $\epsilon \rightarrow 
R \epsilon R^{{\rm T}}$, with $R$ the standard $3 \times 3$ rotation
matrix.

GW's with a given wavevector $\kV$ are produced during inflation and 
then stretched outside the horizon. The amplitude is constant 
outside the horizon, but once a mode reenters its energy redshifts 
as $a^{-4}$. For the inflationary spectrum the effect of very short 
wavelength modes is negligible, and so we can assume that all modes enter 
during the matter-dominated era, for which the exact time evolution is 
given \cite{AbbWs,me} by a spherical Bessel function, 
$3 j_{1}(k \tau)/(k \tau)$. This time evolution is also correct for
all modes long after matter-radiation equality. Inflation 
produces Gaussian, stochastic perturbations. The Fourier components 
have zero ensemble mean and a covariance 
\begin{equation}
\label{frpres}
<h^{i}(\kV,\tau_{1}) h^{j}(\vec{q},\tau_{2})>=\AT k^{-3} \left[ 
\frac{3 j_{1} (k\tau_{1})} {k \tau_{1}} \right] \left[ 
\frac{3 j_{1}(k\tau_{2})}
{k \tau_{2}} \right] \delta^{3}(\kV+\vec{q})\delta_{ij},
\end{equation}
for the scale-invariant $k^{-3}$ spectrum. Note that we do not 
assume the short-wavelength approximation $h^{i}(\kV,\tau) \propto 
a^{-1}(\tau)~ e^{i k \tau}$. The contribution to $\Omega$ at the 
present (averaged over several periods) is 
\begin{equation}
\label{ompres}
\Omega_{\lambda}=\frac{d\Omega_{\rm GW}}{d \ln k}=\frac{3\pi}{2}\AT 
(k\tO)^{-2}\ ,
\end{equation}
where $\tau_{0}=2 H_{0}^{-1}$ is the present value of $\tau$, and 
throughout we set $H_{0}=75\,{\rm km\,sec^{-1}\,Mpc^{-1}}.$
Normalization to the full CMB quadrupole anisotropy gives $\AT =6 
\times 10^{-11}$. 

Consider a photon emitted from a source toward 
an observer at the origin, with the photon's final direction defined
as (minus) the $z$-axis. We use $r$ to denote values of
the $z$-coordinate (with $z_{S}$ denoting the source redshift, not 
its $z$-coordinate). 
GW's affect the distance-redshift relation, but this effect is 
separate from that
of the angular deflections which we are interested in, and it 
introduces only small 
additional corrections in these quantities \cite{linder}.
We can thus neglect this effect, and assume that the photon path 
obeys $r(\tau)=\tau_{0}-
\tau$. In a flat, matter-dominated universe, $r_{S}=2 H_{0}^{-1} 
[1-(1+z_{S})^{-1/2}].$ 
The components perpendicular to the $z$-axis of the photon 
direction obey \cite{linder}
\begin{equation}
\frac{dx^{i}}{d\tau}(\tau)-\frac{dx^{i}}{d\tau}(\tO)=h_{zi}(\tau)-
h_{zi}(\tO)-\frac{1}{2}\int_{\tau}^{\tO}\nabla_{i}h_{zz}(\tau')\ 
d\tau'\ .
\end{equation}
Integrating this we find, for the perpendicular components of the 
position (with respect to $\xV(\tO)=0$),
\begin{equation}
x^{i}(\tau)=\int_{\tau}^{\tO}\left[\frac{1}{2}(\tau'-\tau)\nabla_{i}
h_{zz}(\tau')+h_{zi}(\tO)-h_{zi}(\tau')\right]d\tau'\ .
\label{thetai}
\end{equation}
We define a (two-component) angle $\beta^{i}=x^{i}(\tau)/r(\tau)$.

In gravitational lensing with a primary thin lens at a distance 
$r_L$ (but no LSS or GW's) the lens equation is (e.g., 
\cite{schneider}) 
\begin{equation}
\vec{\beta}=\vec{\theta}-\vec{\alpha}_{\rm lens} (\vec{\theta})\ , 
\label{1lens}
\end{equation}
where $\vec{\theta}$ is the observed image angle, $\vec{\beta}$ is
the source angle (defined as $\vec{x}_S/r_S$, in terms of the 
perpendicular position of the source), and $\vec{\alpha}_{\rm lens}$ 
is the deflection angle scaled by $r_{LS}/r_S$ (we 
define $r_{LS}=r_S-r_L$). 
In this case, the fiducial $z$-axis is defined to be in the observed
direction of the lens. The distortion of the image of a small 
source is given by the inverse of the Jacobian matrix
\begin{equation}
\frac{\partial \beta^{i}}{\partial \theta^{j}}=\delta^{ij}-
\Psi^{ij}\ ,
\end{equation} 
where $\Psi^{ij}$ is also termed the shear tensor of the lens.

\section{Shear induced by GW's on lensing}

In this section we follow the approach used for LSS in Ref.\ 
\cite{seljak}, 
i.e., we compute some of the same quantities for GW's and compare the
results. As stated in Sec.\ I, we do not need to include a lens 
explicitly, as we now justify.
In the presence of a metric perturbation, but without a primary 
lens, the lens
equation has the form $\vec{\beta}=\vec{\theta}-\vec{\alpha}_{OS}
(\vec{\theta})$, 
where $\vec{\alpha}_{OS}$ results from the accumulated deflection 
between the 
observer and the source. As defined in Sec.\ II, the shear tensor for 
an image at 
$\vec{\theta}$ due to the perturbation is $F^{ij}=\partial 
\alpha_{OS}^{i}(\vec{\theta})/ 
\partial \theta^{j}$. On the other hand, the relative deflection 
at $\vec{\theta}$ between 
two rays separated by a tiny angle $\vec{\gamma}$ is $\vec{\alpha}_
{OS}(\vec{\theta}+\vec{\gamma})-\vec{\alpha}_{OS}(\vec{\theta})$. 
We denote the rms of this quantity by $\sigma_{\Delta\beta}$. We 
average over directions of $\vec{\gamma}$ (which
for this calculation is equivalent to fixing $\vec{\gamma}$ and 
assuming that $F^{ij}$ is 
isotropic) and take $\gamma \rightarrow 0$, obtaining the relation
\footnote{Repeated indices are summed over the x and y directions. 
There is no distinction between upper and lower indices.}
\begin{equation}
\left(\frac{\sigma_{\Delta\beta}}{\gamma}\right)^2=\frac{1}{2}
\left<F^{ij}F_{ij}\right>\ ,
\label{FM}
\end{equation} 
all evaluated at position $\vec{\theta}$. Thus, $\sigma_{\Delta
\beta}/\gamma$ yields an estimate of the magnitude of the shear 
tensor. Indeed, it fully characterises rms
values of $F^{ij}$, since for an isotropic field 
\begin{equation}
\left<F_{ij}F_{kl}\right>=\frac{1}{8} \left[\delta_{ij}\delta_{kl}+
\delta_{ik}\delta_{jl}+
\delta_{il}\delta_{jk}\right] \left<F^{mn}F_{mn}\right>\ .
\end{equation}

\begin{figure}[t]
\vspace*{6.3 cm}
\caption{Sketch showing positions of the observer, lens and source,
as well as an image ray and several comoving distances.}
   \includegraphics{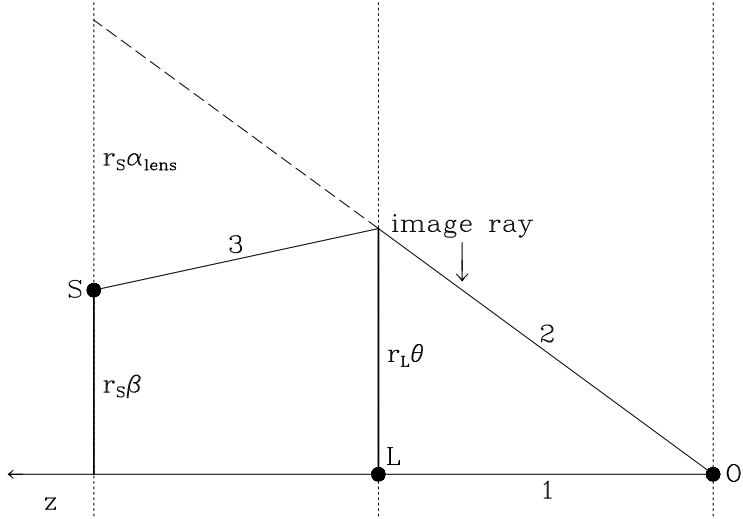}
\label{fig}
\end{figure}

If we also include a primary lens, in the lens equation we simply 
add up all deflections
linearly, assuming all deflections are small. For the primary lens 
alone, we have
Eq.\ (\ref{1lens}). Figure \ref{fig} shows this setup 
schematically. In the
presence of a metric perturbation, we trace a light ray that is 
observed at $r=0$ to
come from the direction $\vec{\theta}$, back to the source.
We find a different form for the lens equation:
\begin{equation}
\vec{\beta}=\vec{\theta}-\vec{\alpha}_{OS}^{(2,3)}-\vec{\alpha}_
{\rm lens} (\vec{\theta}-\vec{\alpha}_{OL}^{(2)})\ . 
\label{2lens}
\end{equation}
Here $\vec{\alpha}_{OS}^{(2,3)}$ refers to the integrated deflection
caused by LSS or GW's along the paths labeled 2 and 3 in Figure 
\ref{fig}, defined so that the total induced
change in $\vec{x}(\tau_S)$ equals $r_S \vec{\alpha}_{OS}^{(2,3)}$. 
Similarly, $r_L \vec{\alpha}_{OL}^{(2)}$ is the induced change in 
$\vec{x}(\tau_L)$. In integrating
the deflections along the unperturbed paths 2 and 3, we are assuming
that the relative deflections due to LSS or GW's are small compared 
to $\theta$ and $\alpha_{\rm lens}$,
which is true for the cases which we consider below. 

When the perturbation is included, $\vec{\theta}$ is no longer an 
observable, since it is measured 
with respect to the unperturbed position of the lens. The observed 
position of the lens (whose actual position has not changed) is now 
$\vec{\theta}_{\rm lens}=\vec{\alpha}_{OL}^{(1)}$, so the lens 
equation in terms of the observable $\vec{\theta}'=\vec{\theta}-
\vec{\theta}_{\rm lens}$ is
\begin{equation}
\vec{\beta}=\vec{\theta}'+\vec{\alpha}_{OL}^{(1)}-
\vec{\alpha}_{OS}^{(2,3)}-\vec{\alpha}_{\rm lens}(\vec{\theta}'+
\vec{\alpha}_{OL}^{(1)}-\vec{\alpha}_{OL}^{(2)})\ . 
\label{3lens}
\end{equation}
If we now calculate the shear tensor resulting from equation 
(\ref{3lens}), it will contain 
the shear of the primary lens, shear terms from the perturbation, 
and also cross terms. 
For simplicity, in the case of GW's we only estimate one 
characteristic
magnitude, that of the shear resulting from $\vec{\alpha}_{OS}^
{(2,3)}$, by evaluating
the corresponding $\sigma_{\Delta\beta}/\gamma$. In Ref.\ 
\cite{me2} all the different shear
terms were studied for LSS, showing that $\sigma_{\Delta\beta}/\gamma$
indeed estimates the relative magnitude of the various corrections
due to LSS. Since we find that $\sigma_{\Delta
\beta}/\gamma$ is much smaller for GW's, we do not have any 
motive to explore equation (\ref{3lens}) further in this case. 
Instead of the path $(2,3)$, we may use a straight path 
from $r=0$ to $r=r_S$ to evaluate the rms of various quantities 
in this section, since $\alpha_{\rm lens} \ll 1$ and so the 
components of vectors and tensors as well as the relative distances
of points along the path (both of which enter 
into the rms calculations) are unchanged (except for $O(\alpha_{\rm
 lens})$ corrections). Thus we only need to consider the effect of 
GW's in the absence of a primary lens. 

Consider first a single light ray with $\vec{\theta}=0$. In the 
absence of GW's (or LSS) it
would follow the straight line $x^{i}(\tau)=0$ for all 
$\tau$. We now include the effect of GW's, and compute the rms 
fluctuation in the 
photon's perpendicular displacement at the source, $\sigma_{\beta}=
\left<\vec{\beta}(\tS)\cdot\vec{\beta}(\tS)\right>
^{1/2}$. This is a measure of the common deflection of all image 
rays, and is therefore
not observable, but it is useful for the calculations that follow.
We use equation (\ref{thetai}) and convert the expression to Fourier
space. Consider first only the $h_{zz}$ term, whose contribution 
to $\sigma_{\beta}^{2}$ we denote $\sigma_{\beta,a}^{2}$. The
polarization gives $(\epsilon_{zz}^+)^2+(\epsilon_{zz}^{\times})^2=
\sin^4\theta_k$, where $\kV=(k,\theta_k,\phi_k)$ in spherical
coordinates. Performing the angular $\kV$ integrations then yields 
$$\sigma_{\beta,a}^{2}=
\frac{432 \pi} {r_{S}^{2}} \AT \int_{\tS}^{\tO}d\tau_{1} \int_{\tS}
^{\tO}d\tau_{2}\int_{0}^{\infty} dk\ k(\tau_{1}-\tS)(\tau_{2}-\tS)
\frac{j_{1}(k\tau_{1})}{k\tau_{1}} \frac{j_{1}(k\tau_{2})}
{k\tau_{2}}\frac{j_{3}[k(\tau_{1}-
\tau_{2})]}{[k(\tau_{1}-\tau_{2})]^3}\ .$$ 
The $j_{3}[k(\tau_{1}-\tau_{2})]/[k(\tau_{1}-\tau_{2})]^3$ term 
represents a
further suppression of short wavelength modes due to phase 
cancellations among
different waves in the assumed isotropic stochastic background. 
Letting $s=k\tau_{1}$ 
and $q=\tau_{2}/\tau_{1}$, we can simplify this expression to a 
double integral,
\begin{equation}
\sigma_{\beta,a}^{2}=\frac{864 \pi} {r_{S}^{2}} \AT 
\int_{\oS}^{1} \left\{q\tO\left(\frac{1}{2}\tO-\tS\right)
-\tS r_{S}+\tS^{2} \left[\frac{1}
{2q}+\ln(q/\oS)\right]\right\}\,W(q)dq\ ,
\end{equation}
where $\oS=\tS/\tO$ and with $u \equiv (1-q)s$ we define
\begin{equation}
W(q)=\int_{0}^{\infty}\frac{j_{1}(s)}{s}\frac{j_{1}(q s)}{q s}
\frac{j_{3}(u)}{u^3}\,s\ ds\ .
\end{equation}
Similarly, the contribution of the $h_{zi}$ terms of Eq.\ 
(\ref{thetai}) is
\begin{equation}
\sigma_{\beta,b}^{2}=\frac{144 \pi}{r_{S}^{2}} \AT \int_{\oS}^{1} 
\left[\frac{\tS^{2}}{2 q^{2}}+\tO\left(\frac{1}{2}\tO-\tS\right)
\right]G(q)dq\ ,
\end{equation}
where
\begin{equation}
G(q)=\int_{0}^{\infty}\left\{\frac{1}{5} \left[\frac{j_{1}(s)}
{s}\right]^{2}+\frac{1}{5}\left[\frac{j_{1}(q s)}{q s}\right]
^{2}-2\,\frac{j_{1}(s)}{s}\frac{j_{1}(q s)}{q s}\left[\frac
{j_1(u)}{u}-2\frac{j_2(u)}{u^2}\right]
\right\}\frac{ds}{s}\ .
\end{equation}
Integrating over angles gives a zero cross term, and so $\sigma_
{\beta}^{2}=\sigma_{\beta,a}^{2}+\sigma_{\beta,b}^{2}$.
Numerically, we find that $\sigma_{\beta}=5 \times 10^{-6}\ 
(z_{S}=1),\ 9 \times 10^{-6}\ (z_{S}=3)$. This is much smaller 
than the estimates for LSS \cite{seljak}, $6 \times 10^{-4}\ 
(z_{S}=1),\ 7 \times 10^{-4}\ (z_{S}=3)$. 

To estimate the relative deflection between rays at $\vec{\theta}
=0$, we choose two 
directions (labeled A and B) separated at the observer by an 
infinitesimal angle $\gamma$, 
and find the rms difference between the deflections due to GW's in 
these two directions, 
$\sigma_{\Delta\beta}=\left<\left[\vec{\beta}_{A}(\tS)-
\vec{\beta}_{B}(\tS)\right]^{2}\right>^{1/2}$.
We cannot evaluate this with the method used for LSS, which assumes 
that horizon size modes are negligible \cite{kaiser}. Instead we 
must calculate $\sigma_{\Delta\beta}^{2}$ explicitly and keep all 
the terms to lowest order in $\gamma$, i.e., quadratic order. These 
include terms which come from multiplying polarization components 
for the different directions A and B. The final result is
\begin{equation}
(\sigma_{\Delta\beta}/\gamma)^{2}=4 \sigma_{\beta,a}^{2}+
\sigma_{\Delta\beta,a}^{2}+\sigma_{\Delta\beta,b}^{2}
+\sigma_{\Delta\beta,c}^{2}\ ,
\end{equation}
where
\begin{eqnarray}
\sigma_{\Delta\beta,a}^{2}&=&\frac{576 \pi}{r_{S}^{2}} \AT 
\int_{\oS}^{1}\left[\frac{\tO\tS}{q}(1+q)^2-\left(\tO^2+\frac{
\tS^2}{q}\right)(1+q)+(\tO+\tS)r_S \ln(q/q_S)\right] I_1(q)dq\ ,
\nonumber \\
\sigma_{\Delta\beta,b}^{2}&=&\frac{1728 \pi}{r_{S}^{2}} \AT 
\int_{\oS}^{1}
\biggl\{(\tO+\tS)^2-\tO^2(1+q)^2+2q\tS(\tS-q\tO)+ \\ \nonumber
& &\hspace{1in}+\left[\tS\tO(1+q^2)+q(\tO+\tS)^2\right]
\ln(q/\oS)\biggl\}I_2(q)dq\ , \nonumber \\
\sigma_{\Delta\beta,c}^{2}&=&\frac{288 \pi}{r_{S}^{2}} \AT
\int_{\oS}^{1} \left[\frac{\tS^{2}}{2 q^{2}}+\tO\left(
\frac{1}{2}\tO-\tS\right)\right] I_3(q) dq\ , \nonumber \\
I_1(q)&=&\int_{0}^{\infty}\frac{j_{1}(s)}{s}\frac{j_{1}(q s)}{q s}
\left[\frac{j_{2}(u)}{u^2}-3\frac{j_{3}(u)}
{u^3}\right]s\,ds\ , \quad 
I_2(q)=\int_{0}^{\infty}\frac{j_{1}(s)}{s}\frac{j_{1}(q s)}{q s}
\frac{j_{4}(u)}{u^4}s^3\,ds\ , \nonumber \\
I_3(q)&=&\int_{0}^{\infty}\left\{\frac{2}{15}\left[\frac{j_{1}(s)}
{s}\right]^{2}+\frac{2}{15}\left[\frac{j_{1}(q s)}{q s}\right]
^{2}-2\,\frac{j_{1}(s)}{s}\frac{j_{1}(q s)}{q s}\left[\frac
{j_1(u)}{u}-3\frac{j_2(u)}{u^2}\right]
\right\}\frac{ds}{s}\ . \nonumber 
\end{eqnarray}
Then $\sigma_{\Delta\beta}/\gamma=7 \times 10^{-6}\ (z_{S}=1),\ 1.3 
\times 10^{-5}\ 
(z_{S}=3)$. By contrast, LSS gives a $\sigma_{\Delta\beta}/\gamma=0.
07\ (z_{S}=1),\ 0.14\ (z_{S}=3)$. For LSS, the relative deflection 
is greatly increased by coherent deflections for short wavelength 
modes, but for GW's the effect of these modes is cut off 
by the redshifting as well as the temporal oscillations. We also 
used the relation 
$k r_{S}\gamma \ll 1$ in the calculation of $\sigma_{\Delta\beta}$.
The reason we find a $\sigma_{\Delta\beta}$ of order 
$\gamma\sigma_{\beta}$ is that long wavelength modes overlap over 
the two light rays, and the relative deflection is small compared 
to the total deflection. Indeed, a Taylor expansion suggests that 
in general $\sigma_{\Delta\beta}/\gamma \sim k
 r_{S} \sigma_{\beta}$, and $k r_{S} \sim 1$ is dominant for this 
GW spectrum. As shown above, the shear tensor (which is also used 
in weak lensing) is closely related to $\sigma_{\Delta\beta}/
\gamma$, and so the mean square ellipticity at a point induced by GW's 
is of order $10^{-5}$, again negligible compared to the few per 
cent expected from LSS (e.g., \cite{kaiser}).

We can also try to derive general limits on GW's at astrophysical 
wavelengths from the induced shear. To obtain a limit on 
$\Omega_{\lambda}$, we compute the $\sigma_{\Delta\beta}/\gamma$ 
produced by an isotropic background of GW's at a single 
wavenumber $k$. Note that for 
modes at a given $k$, we can use equations (\ref{frpres}) and 
(\ref{ompres}) even for short wavelengths (with $\AT$ a 
normalization factor, separate for each $k$), for times $\tau$ 
long after matter-radiation equality. Since GW's at Horizon wavelengths 
are already strongly constrained by the CMB as noted above, we 
restrict our calculation to the case $k r_S \gg 1$, in which case 
the $h_{zi}$ terms in Eq.\ (\ref{thetai}) can be neglected. 
We can estimate from Eq.\ (\ref{thetai}) that in order of 
magnitude $\sigma^2_{\beta}$ should equal $\AT/(k \tO)^4$, and thus 
that $\sigma^2_{\Delta\beta}/\gamma^2 \sim \AT/(k \tO)^2$. However,
we find from the exact calculation that there is no term this
large, only higher order terms in $1/(k \tO)$. We outline in 
the Appendix a mathematical argument showing this cancellation 
at small wavelengths. This result requires both the phase 
cancellations that come in averaging over an isotropic 
background, and also the oscillation with time of the GW's.
With different assumptions, e.g., if we analyzed GW's from a 
particular source, which are then not isotropic, 
stronger limits may be possible.

\section{Proper motions induced by LSS and GW's}

We now consider the fluctuation of the angular deflection of image 
rays with time, and the 
resulting proper motion. If the deflection of image rays induced by
 LSS or GW's changes
significantly during an observation of a lens system, then the slow
 shift in alignment
between the lens and the source will change the impact parameter at
the lens of a given ray from the source. The images will therefore 
move, and even tiny motions
may be detected since the source motion is magnified if it is 
lensed by a primary lens.
We first show that this effect is still expected to be too small to
 measure for LSS and for 
the GW power spectrum that we have considered above. However, given
 the weakness of existing 
limits on GW's at astrophysical wavelengths (Sec.\ I), we also consider 
possible limits on a general GW spectrum.

Again we consider a single light ray from the observer out to some 
distance $r_S$, in the absence of a primary lens (we consider the 
effect of a lens below). Given a ray with a fixed direction at the 
observer, its position $x^{i}(r_{S})$ at $r_S$ moves with time, and
it is this motion which we evaluate. In practice, we 
are interested in a fixed source at $r_S$, in which case its {\it 
apparent} position will drift with the same speed but in the 
opposite direction. For LSS we have (e.g., \cite{seljak})
\begin{equation}
x^{i}(r_{S})=-2\int_{r=0}^{r_{S}}(r_{S}-r)\nabla_{i}\phi
(\tau=\tO-r)dr,
\end{equation}
in terms of the Newtonian potential (or scalar metric perturbation) 
$\phi$. We are now using the parameter $r$ rather than $\tau$, 
since as time changes all comoving distances remain fixed. The only 
change is the time of evaluation of $\phi$, and so
to find $dx^{i}(r_{S})/d\tO$ from $x^{i}(r_{S})$ we simply replace 
$\phi(\tau=\tO-r)$ by $\dot{\phi}(\tau=\tO-r)$, with the partial 
time derivative in $\dot{\phi}$ taken at a fixed position. The rms 
value of $dx^{i}(r_{S})/d\tO$ depends on the power spectrum of 
$\dot{\phi}$, a quantity which has been estimated by various 
authors in connection with the Rees-Sciama effect on the CMB 
(e.g., \cite{RS}). While the integrated deflection is dominated by 
short ($\sim 1$ Mpc) wavelengths, the LSS potential only evolves 
on a cosmological time scale. In an Einstein-deSitter universe, 
$\phi$ is time independent in the linear regime of small density 
perturbations, but in this case too $\dot{\phi}$ becomes nonzero 
when nonlinear structure forms. In general, therefore, the 
proper motion induced by LSS is of order $\sigma_{\beta}/\tO \simeq 
10^{-8}$ arcsec per year. For the gravity wave spectrum considered 
above, horizon size modes are dominant, and so here too the induced 
proper motion is of order $\sigma_{\beta}/\tO$, with a $\sigma_
{\beta}$ smaller by $\sim 10^{2}$ than for LSS. Any observed proper 
motion will thus be dominated by peculiar velocities of hundreds 
of km/s generated, e.g., by the velocity dispersion of stars in a 
galaxy or galaxies in a galaxy group or cluster.

We now estimate the lensing limit on stochastic GW's in general, at 
any wave number $k$. VLBI observations can directly measure or 
limit proper motions, and this then implies a limit on GW's. 
Again we restrict ourselves to wavelengths with $k r_S \gg 1$, and 
consider first the apparent motion of a source that is not lensed by a 
primary lens. The apparent motion due to GW's 
of a fixed object at distance $r_S$ is $-d\vec{\beta}(r_S)/
d\tau_0$. Up to corrections of order $1/k r_S$, the mean 
square of this motion is
\begin{equation}
\left<\left[\frac{d}{d\tau_0}\beta(r_S)\right]^2\right> = \frac{18
 \pi \AT} {5 k^2 \tO^4} \left[1+\frac{2}{3}\cos(2 k \tO)\right]\ .
\label{eqcom}
\end{equation}

However, when there are both a lens and a source, a GW background 
will produce correlated proper motions in both, and the relative
motion may be small. Limits from VLBI on proper motions in 
gravitational lenses were recently considered
in Ref.\ \cite{koch}, and we proceed similarly. We may
hope for strong limits because, in the presence of lensing, a 
proper motion of the source relative to the lens is magnified into 
a larger proper motion of the images. 
Furthermore, only a {\it relative} motion between images needs to 
be detected, as opposed to a more difficult measurement of motion 
with respect to an external reference frame, since if the source moves 
(relative to the lens), the different images do {\it not} all move 
together. In general, different values of the magnification matrix 
at the different image positions will produce relative motions 
between images of the same order of magnitude as the absolute 
motions. Moreover, pairs of highly magnified 
images generally have antiparallel motions \cite{koch}. 

To analyze how proper motion due to GW's may be magnified, we
start from Eq.\ (\ref{3lens}), and consider the same equation a 
time $\Delta t$ later, when the deflections from GW's have changed. 
E.g., $\vec{\alpha}_{OL}^{(1)}$ has changed to $\vec{\alpha}_{OL}^
{(1)}+\Delta^{(1)}$, and a total change $\vec{\Delta}$ in the 
observed $\vec{\theta}'$ has been induced. Expanding the lens 
equation to first order in the small changes and solving for 
$\vec{\Delta}$, we obtain 
\begin{equation}
\Delta_{i}=\Delta^{(2)}_{i}-\Delta^{(1)}_{i}+M_i^j \left[\Delta^
{(2,3)}_j-\Delta^{(2)}_j\right]\ ,
\label{eqdel}
\end{equation}
where the magnification matrix $M_i^j$ equals the inverse of 
$\delta_i^j-\partial_i\alpha_{\rm lens}^j$ (and is evaluated at 
$\vec{\theta}'+\vec{\alpha}_{OL}^{(1)}-\vec{\alpha}_{OL}^{(2)}$). 
Consider first the magnified term, $\Delta^{(2,3)}_j-\Delta^{(2)}_j$. 
Averaging over directions of $\vec{\Delta}^{(2,3)}-\vec{\Delta}^{(2)}$ 
we obtain a result analogous to Eq.\ (\ref{FM}) for the mean 
square. Since $M^{ij}$ is symmetric for a thin lens \cite{schneider}, 
it has two real eigenvalues $m_a$ and $m_b$ (where the magnification 
$M=|m_a m_b|$). Letting 
\begin{equation}
\Mtil=\left[\frac{1}{2}(m_a^2+m_b^2)\right]^{1/2}\ ,
\label{Mtil}
\end{equation}
we find that
\begin{equation}
{\rm rms\ }\left|M_i^j \left[\Delta^{(2,3)}_j-\Delta^{(2)}_j\right]
\right| = \Mtil\, \times 
{\rm rms\ }\left|\vec{\Delta}^{(2,3)}-\vec{\Delta}^{(2)}\right|\ . 
\label{eq2mag}
\end{equation}
In Eq.\ (\ref{eq2mag}) we may evaluate the rms on the right-hand 
side using a straight path (as in Sec.\ III). 
Letting
\begin{equation}
\frac{d}{d\tO}\vec{\beta}_{LS} =-\frac{d}{d\tO}\vec{\beta}(r_S)+
\frac{d}{d\tO}\vec{\beta}(r_L)\ ,
\end{equation}
we find that in $\left<\dot{\beta}_{LS}^2\right>$ there is no
term of order $\AT /(k^2 \tO^4)$ [as in Eq.\ (\ref{eqcom})],
but only higher order terms in $1/(k \tO)$.
Once again this small wavelength cutoff results from combining 
the time oscillation of GW's and the phase cancellations in averaging 
over an isotropic background (see the Appendix), and as a result
there is only a very weak limit on $\Omega_{\lambda}$.

\section{Conclusions}

Gravitational lensing is affected by perturbations to the 
homogeneous and isotropic
background metric. Such perturbations, whether they are caused by 
LSS or GW's, may produce a number of effects on light propagation. 
One such effect is an overall shift in the angular positions of 
nearby objects, which is not observable. Another is 
a relative difference between the induced shifts in nearby light 
rays. This relative deflection manifests itself as a shear which 
may cause weak lensing
and also affect strong lensing. A third effect is a fluctuation of 
the angular position of distant objects with time, leading to a 
directly observable proper motion. 

The actual amplitude of long wavelength modes of LSS and GW's is 
limited by the quadrupole anisotropy of the CMB. Even if both make 
comparable contributions to the anisotropy, LSS is dominant in its 
effects on lensing. This results from cancellations due to the time 
oscillation of short wavelength gravity waves, as well as the 
redshifting of their amplitude.
For LSS, on the other hand, the effect of small coherent structures
is amplified as the deflection executes a random walk. We find that 
the relative deflection due to GW's is four orders of magnitude 
smaller than that of LSS, and is therefore not observable. 

The induced proper motions expected for LSS or for GW's generated in 
inflation are small compared to typical peculiar velocities, and 
thus are not observable. The motions are also too small to yield
interesting limits on the energy density of GW's at shorter 
wavelengths.

After this paper was submitted for publication, the bending of
light by gravity waves was analyzed differently in Ref.\ 
\cite{new}, for the case of short (subhorizon) wavelengths,
in a nonexpanding flat space (i.e., neglecting the redshifting
of the amplitude of GW's). That simplified analysis shows that the
relative proper motion between two sources is small not only if
they are at different redshifts along the same line of sight
(in agreement with our calculation of 
$\left<\dot{\beta}_{LS}^2\right>$ in Sec.\ IV), but also
if they are separated on the sky by a small angle. The
treatment presented in Ref.\ \cite{new} changes quantitatively
if expansion is included, but not qualitatively for GW's with
a period short compared to the redshifting time scale (i.e.,
a Hubble time).

\vspace{.1in}
\begin{center}
{\bf Acknowledgements}
\end{center}

It is a pleasure to thank Ed Bertschinger for helpful advice and many 
valuable comments, Uro\v{s} Seljak for valuable discussions and 
comments, and Paul Schechter, Sean Carroll, Alan Guth and Paul 
Steinhardt for helpful discussions. I thank Scott 
Dodelson for pointing out an error. This work was supported by NASA 
grant No.\ NAG5-2816.

\appendix
\section{Appendix}

In trying to set limits on GW's at short wavelengths $k r_S \gg 1$,
we twice encountered a weaker limit than simple dimensional
analysis would suggest: Once in calculating 
$\sigma_{\Delta\beta}/\gamma$ or shear in Sec.\ III, and then in
estimating the magnified proper motion in Sec.\ IV. In this 
Appendix we outline the first of these calculations and
show how this result emerges. The second calculation can be 
done similarly. 

From Eq.\ (\ref{thetai}) in the limit of short wavelengths
(compared with the present Horizon), we derive
$$\sigma_{\Delta\beta}^{2}/\gamma^2=
\frac{1728 \pi} {r_{S}^{2}} \AT k^4 \int_{\tS}^{\tO}d\tau_{1} 
\int_{\tS}^{\tO}d\tau_{2}(\tau_{1}-\tS)(\tau_{2}-\tS)
(\tO-\tau_{1})(\tO-\tau_{2})
\frac{\cos(k\tau_{1})}{(k\tau_{1})^2}\frac{\cos(k\tau_{2})}
{(k\tau_{2})^2}\frac{j_{4}[k(\tau_{1}-
\tau_{2})]}{[k(\tau_{1}-\tau_{2})]^4}\ .$$ 
The $j_{4}[k(\tau_{1}-\tau_{2})]/[k(\tau_{1}-\tau_{2})]^4$
term comes from the angular $\kV$ integrations, including
the angular dependence of the polarizations and assuming
an isotropic background. 
Letting $x=k \tau_1$ and $u=k (\tau_1-\tau_2)$ (also $x_0
=k \tO$, etc.) leads to
$$\frac{864 \pi} {k^2 r_S^2} \AT \int_{x_S}^{x_0}dx
\frac{(x-x_S)(x_0-x)}{x^2}\int_{x_S-x}^{x_0-x}du\frac{
(x+u-x_S)(x_0-x-u)}{(x+u)^2} \frac{j_4(u)}{u^4}\left[
\cos u(1+\cos 2 x)-\sin u \sin 2 x\right]\ . $$ 
We evaluate only the first $\cos u$ term here, since the other 
terms in the square brackets can be evaluated similarly. Note
that dimensional analysis suggests that the $x$ and $u$ integrals
should give a term of order 1 (not larger, because of the
oscillating integrand).

To do the $u$ integral we separate the smooth and oscillating 
parts and then repeatedly integrate by parts: We let $w^{[n]}(u)$ 
be the $n$-th indefinite integral of $[\cos u]\,j_4(u)/u^4$ with respect to 
$u$, and $v^{[n]}(u)$ the $n$-th derivative of $(x+u-x_S)
(x_0-x-u)/(x+u)^2$ with respect to $u$. For each $n$ such that $w^{[n]}(u)$
converges for $u\rightarrow\pm\infty$, we fix the arbitrary constant by
$w^{[n]}(\infty)+w^{[n]}(-\infty)=0$ (Any constant will do 
for other $n$). Then the $u$ integral equals a 
series of terms evaluated at the two limits of integration,
$$ \sum_{n=0}^{\infty}(-1)^n\left\{ w^{[n+1]}v^{[n]}|_{u=x_0-x}-
w^{[n+1]}v^{[n]}|_{u=x_S-x}\right\}\ . $$
Since the two series of terms can be handled similarly, we 
evaluate here only the $u=x_0-x$ terms. We do the $x$ integration
in the same way as the $u$ integration. Thus we continue to 
integrate $w^{[n+1]}(x)$ with respect to $x$, and let $v^{[n,m]}(x)$
be the $m$-th derivative of $(x-x_S)(x_0-x)v^{[n]}(x_0-x)/x^2$
with respect to x. The contribution to $\sigma_{\Delta\beta}^{2}/
\gamma^2$ from the terms we have kept is
$$\frac{864 \pi} {x_S^2} \AT \sum_{n,m=0}^{\infty} (-1)^{n+1}
\left\{w^{[n+m+2]}(0)v^{[n,m]}(x_0)-w^{[n+m+2]}(x_0-x_S)
v^{[n,m]}(x_S)\right\}\ . $$ 
Now, $v^{[n,m]}(x)$ at $x \gg 1$ is of order $x^{-(n+m)}$,
$w^{[n]}(0)$ is $0$ or a constant, and we find that 
$w^{[n]}(x)$ at $|x| \gg 1$ is of order $|x|^{n-5}$. This last
fact, that $w^{[n]}(\pm\infty)$ converges for the first few
$n$, depends on the specific function $w^{[0]}(x)$ which 
in turn is determined by the two physical assumptions of 
time oscillation and angular averaging. The only term
from the final sum that could give a contribution of order
$\AT/x_S^2$ is the $n=m=0$ term. We find that $w^{[2]}(0)$
is a nonzero constant, but since $v^{[0,0]}(x)=0$ identically,
there is no term of this lowest order.

\end{document}